\documentstyle[11pt,epsf]{article}
\newcommand{\be}{\begin{equation}}
\newcommand{\ee}{\end{equation}}
\newcommand{\benn}{\begin{equation*}}
\newcommand{\eenn}{\end{equation*}}
\newcommand{\bea}{\begin{eqnarray}}
\newcommand{\eea}{\end{eqnarray}}

\def\beq{\begin{eqnarray}}
\def\eeq{\end{eqnarray}}
\def\bea{\begin{eqnarray*}}
\def\eea{\end{eqnarray*}}

\def\centeron#1#2{{\setbox0=\hbox{#1}\setbox1=\hbox{#2}\ifdim
\wd1>\wd0\kern.5\wd1\kern-.5\wd0\fi
\copy0\kern-.5\wd0\kern-.5\wd1\copy1\ifdim\wd0>\wd1
\kern.5\wd0\kern-.5\wd1\fi}}
\def\ltap{\;\centeron{\raise.35ex\hbox{$<$}}{\lower.65ex\hbox{$\sim$}}\;}
\def\gtap{\;\centeron{\raise.35ex\hbox{$>$}}{\lower.65ex\hbox{$\sim$}}\;}

\def\singleandthirdspaced{\baselineskip=\normalbaselineskip\multiply
    \baselineskip by 130\divide\baselineskip by 100}
\def\singlespaced{\baselineskip=\normalbaselineskip}

\newcommand{\newc}{\newcommand}
\newc{\qbar}{{\overline q}}
\newc{\Kahler}{K\"ahler }
\newc{\deltaGS}{\delta_{\rm GS}}
\begin{document}
\begin{titlepage}
\begin{flushright}
{\large hep-th/0303252 \\ SCIPP-2003/03\\
}
\end{flushright}

\vskip 1.2cm

\begin{center}

{\LARGE\bf On the Possibility of Large Axion Decay Constants}

\vskip 1.4cm

{\large  Tom Banks, Michael Dine, P. J. Fox, E. Gorbatov }
\\
\vskip 0.4cm
{\it Santa Cruz Institute for Particle Physics,
     Santa Cruz CA 95064  } \\

\vskip 4pt

\vskip 1.5cm

\begin{abstract}
The decay constant of the QCD axion is required by observation to
be small compared to the Planck scale. In theories of ``natural
inflation," and certain proposed anthropic solutions of the
cosmological constant problem, it would be interesting to obtain a {\it
large} decay constant for axion-like fields
from microscopic physics. String theory is
the only context in which one can sensibly address this question.
Here we survey a number of periodic fields in string
theory in a variety of string vacua.
In some examples, the decay constant {\it can} be parameterically
larger than the Planck scale
but the effective action then contains appreciable harmonics of order
$f_A/M_p$.  As a result, these fields are no better inflaton candidates
than Planck scale axions.

\end{abstract}

\end{center}

\vskip 1.0 cm

\end{titlepage}
\setcounter{footnote}{0} \setcounter{page}{2}
\setcounter{section}{0} \setcounter{subsection}{0}
\setcounter{subsubsection}{0}

\singleandthirdspaced

\section{Introduction}

It is probably fair to say that there still does not exist a
compelling microscopic theory of inflation.  In conventional field
theory models of slow roll inflation, it is difficult to
understand the required flatness of the potential.  In chaotic
models, one has similar issues, particularly since one is
interested in fields very large compared to the Planck scale,
where conventional effective field theory notions are highly
suspect.

It has long seemed plausible that string moduli might play a role
in inflation\cite{gaillard}.  One of us\cite{banks} (see also
\cite{thomas}) has described rather detailed scenarios.  In these,
one supposes that there is an approximate modulus, with a
potential whose magnitude is of order $\sqrt{M_{gut}^3 / M_p}$,
scale of variation $M_p$, and has a minimum with vanishing
cosmological constant. Fields with such features are natural in
the Horava-Witten limit of string theory\cite{hw}, or other
brane-world scenarios with higher dimensional Planck scale of
order the GUT scale.  In this picture, fluctuations are
automatically of the correct order. One obtains slow roll and an
adequate number of e-foldings provided that certain parameters
which are naturally of order one are instead of order $10-100$.

There is another widely studied model for inflation, ``natural
inflation", in which pseudo-Goldstone bosons play the role of
inflatons\cite{natural}.  In this paper, we will refer to such
periodic fields generically as ``axions."  For this to work, it is
necessary that the decay constants of these Goldstone bosons,
$f_A$, be large compared to $M_p$, the four-dimensional Planck
scale.  This possibility has been viewed with some skepticism. As
in the case of chaotic inflation, one is in a regime where
effective field theory is suspect.

There is another context in which a large decay constant is
possibly of interest.  As discussed long ago in \cite{bankslight},
and more recently by \cite{villenkin}, if one has an
{\it extremely} light scalar field, with mass smaller than the inverse
Hubble radius today, which can vary over a very large distance in
field space, one can hope to implement an anthropic solution to
the cosmological constant problem.  For example, if, during inflation, there was an
extremely large number of e-foldings,  then there will be regions with
cosmological constants taking a wide range of values; some will be
close to that observed today, and may be selected by anthropic
considerations\cite{anthropiclambda}.  One of us has
argued\cite{sti} that in string theory this is implausible.  For
scalar moduli (i.e. moduli without periodicity) such small masses
(i.e. small compared to the supersymmetry breaking scale) do not
arise in known string models.  This paper also argued that
periodic moduli typically do not have the decay constants required -- many
orders of magnitude larger than the Planck scale.

Arkani-Hamed et. al., hoping to understand natural inflation,
 have recently suggested that one might obtain parameterically
 large $f_A/M_p$
in the context of theories with extra dimensions\cite{nimaetal}, see also
 \cite{Kaplan:2003aj} for a discussion of similar ideas.
They focus on a Wilson line in a fifth dimension, and note that
provided a certain condition on the radius of the the fifth
dimension and the gauge coupling in this dimension is satisfied,
the decay constant can be large.

String theory is replete with periodic moduli.  In addition to
Wilson lines, these arise from higher
rank antisymmetric tensor fields and from inherently stringy
sources.  In this note, we survey
periodic moduli in a variety of string models, asking whether
there are regimes of moduli space where decay constants are
parametrically large.  One might suspect that it is difficult to
make decay constants arbitrarily large. For example, in
compactifications of the heterotic string, the condition of
\cite{nimaetal} requires that the compactification radius be small
in string units. But then there is a $T$-dual description in which
the radii are large, and in which the decay constant must be
small.  This behaviour, we will see, is typical of periodic moduli
in toroidal string and M-theory compactifications.  With enough
supersymmetry, it is a theorem\cite{wittenusc} that extreme
regions in moduli space always can be mapped into large
radius/weak coupling regions of some other string theory (or eleven
dimensional supergravity). We will
also investigate a variety of string compactifications with little
or no supersymmetry, and show that even when a large $f_A / M_p$
can be obtained, the axion potential has variation of order one on
the scale $M_p$, so that large numbers of e-foldings of slow roll
inflation do not occur.  This can be understood as due to the
dominance of higher harmonics ($n \sim f_A/M_p$).   In specific examples,
we will explain how the appearance of additional states and/or
instantons at small radius modifies the naive expectations for the
axion periodicity.

This paper is organized as follows.  In the next section, we discuss string
theories with 32 or 16 supersymmetries.  We will see explicitly for such
theories how the decay constant is bounded by $M_p$, as expected from the
theorem cited above.  We then turn to string configurations with less
supersymmetry.  Here it is difficult to obtain a general result, but in a
variety of examples with eight, four or zero supersymmetries we again find
that the effective decay constant cannot be larger than the Planck scale.  In
section four, we consider similar issues in the eleven dimensional limit. We
present our conclusions in section five.

\section{String Configurations with 16 or 32 Supersymmetries}

In this section, we consider string theories with 16 or more supersymmetries.
We first study compactifications of the heterotic string at weak coupling, since they provide a
simple  realization of the Wilson lines discussed in \cite{nimaetal}.  We then
turn to Type I theories at weak coupling, compactified on tori, and finally
weakly coupled Type II theories.  In all cases, we will see that the decay
constants are bounded by $M_p$.  We explain how a
theorem due to Witten\cite{wittenusc} allows us to understand
these results in a unified way.

\subsection{Weakly Coupled Heterotic Strings}

We first consider a class of theories in which the decay constants
cannot be arbitrarily large.  These are theories which exhibit
some sort of $T$ duality.  In closed string theories compactified
on tori, there are a variety of periodic moduli.  For
definiteness, consider first the weakly coupled heterotic string
theory, and compactify on a six torus whose bosonic action has the form
(here, and throughout this paper, since we are interested in the parametric
dependence of various quantities, we will not be careful about numerical
factors):
\be
S_{bos}= M_s^8\int d^{10}x \sqrt{-g}e^{-2\phi}(R-\frac{1}{2}|H_3|^2)-
M_s^8\int d^{10}x\sqrt{-g}e^{-2\phi}\frac{1}{2}|F_2|^2,
\ee
where $H_3=dB_2$ and $F_2=dA_1$.  For simplicity we will take the torus to
have all radii of size $R$.  The generalization to asymmetric compactifications
is straightforward and does not change our conclusions.  Periodic moduli
include:
\begin{enumerate}
\item  Moduli from $B_{ij}$, the second rank antisymmetric tensor
fields.  Here the axion field is,
\be
\theta = M_s^2 \int d\Sigma^{ij}B_{ij}.
\ee
These have periods of order $1$ in string units.  Dimensionally
reducing the kinetic term for $B_{ij}$ one finds their
decay constants are:
$${f_a^2 \over M_p^2} = {1 \over (R M_s)^4}.$$

\item  Moduli from $B_{\mu \nu}$ (the model-independent axion).
Decay constant\footnote{Some of the estimates we will present, may
be modified by relatively large factors like $16\pi^2$.  We do not
know of examples where such factors help rather than hinder slow
roll inflation.  At any rate, factors like this could at best give
of order $10^2 $ e-foldings.} :
$${f_a^2 \over M_p^2} = 1.$$
\item  Wilson lines (the types of fields considered in
\cite{nimaetal}):  $A_I$ have periods of order $1/R$ in string
units (if one takes the $X^I$ coordinates to have periodicities of
order $R$);
$${f_a^2 \over M_p^2} = {1 \over (R M_s)^2}.$$
\end{enumerate}

Of these, it would appear that the $B_{IJ}$ and $A_I$ moduli can
have large decay constants if $R$ is small.  One can check that it
is precisely this small $R$ region which is the region of moduli
space considered in \cite{nimaetal}.  But because of $T$-duality, one
knows that there is an equivalent description in terms of a large
radius theory, and from our formula, it appears that decay
constants are never large in the large $R$ region.

To see how $T$ duality works it is helpful to examine the exact
spectrum at tree level.  For the case of a single compact
dimension (the generalization to more dimensions is
straightforward), the left-moving momenta in the compact dimension
are:
\beq
K_{L 5}= {n \over R} + {w R \over \alpha^\prime} -q^I A_5^I - {wR
\over 2} A_5^I A_5^I.
\label{leftmomenta}
\eeq
Here the $q_I$'s lie in the momentum lattice.  There are similar
expressions for the other momenta.  The periodicity
$A_5^I \rightarrow A_5^I + {1 \over R}$ is manifest in this
expression.  For small $R$, however, eqn. \ref{leftmomenta}
exhibits an
additional, approximate periodicity,
\beq
A_5^I \rightarrow A_5^I + \frac{R}{\alpha^\prime}.
\eeq
This is as expected for small radius.  This symmetry appears to be only
approximate.  This is because the $T$-duality transformation of
$A_5$ is somewhat complicated.  For small $A_5$, one has simply
$A_5 \rightarrow A_5^\prime=A_5$, but there are non-linear modifications of
this transformation\cite{Giveon:1994fu}.
Note that this means that if we call the periodic variable
$\theta$, in the large $R$ regime,
$\theta = A_5 R$, but in the small $R$, $T$-dual regime,
\beq
\theta =
A_5^\prime
\tilde R \approx {A_5 \alpha^\prime \over R}.
\eeq
So in the large $R$ regime, the decay constant is $1 \over R^2
M_s^2$ is Planck units, but in the small $R$ regime it is $R^2
M_s^2$.

If one breaks supersymmetry, e.g. with Scherk-Schwarz boundary
conditions\cite{rohm}
one can also see this connection at the level of the potential.
For large $R$, the potential receives its largest contribution
from the momentum states; the contributions of winding states are
exponentially suppressed.  In the small $R$ region, the role of
the two sets of states are reversed, and the periodicity of the
dominant terms is different in the two cases.

It is instructive to ask how all of this looks if one works in
terms of the original variables, with the $1/R$ periodicity.  In
this case, for small $R$, the actual periodicity of the potential
(and other quantities in the effective action) is very large, but
the approximate periodicity is small.   This means that in the
Fourier expansion in  $\cos(n\theta A_5 R)$, the typical $n$'s are
large, $n \sim {\alpha^\prime /R^2}$.  So the potential varies,
not over a range $R$, but rather over $\alpha^\prime/R$.  Later we will see a
similar phenomenon in examples where there are no exact dualities
relating the singular limit to a large smooth manifold.

One can consider the other periodic fields similarly.  For the
antisymmetric tensor $B_{IJ} = \phi(x) $, with $I,J$ denoting
coordinates internal to the
torus, the kinetic terms have the form
$$G_N {\alpha^{\prime 2} \over R^4 }(\partial \phi)^2.$$
So
$${f_A^2 \over M_p^2} =  {\alpha^{\prime 2} \over R^4 }.$$
Again, this expression becomes large, formally, for small $R$.
But in a T-dual picture, it has the same form, and so must be
small.  One can trace this back, again, to the details of the
$T$-duality transformation, or to the slow convergence of the
Fourier series in the original variables.


\subsection{Type I Theories}

Consider Type I string theory compactified on a six torus.  We will
be interested in letting the size of one dimension become large or small, so
we will take the volume to be $V_5 R$,
where the direction $X_1$ is of size $R$.  The low energy effective action is
that of type I supergravity whose bosonic action contains
\be
S_{bos}= M_s^8\int d^{10}x\sqrt{-g}e^{-2\phi}R-M_s^8 \int d^{10}x
\sqrt{-g} e^{-\phi}Tr |F_2|^2 + \dots.
\ee
The fields arising from the NS-NS sector
behave similarly under $T$-duality to the antisymmetric tensor
fields of the heterotic string we have described above and do not result in
decay constants larger than $M_p$.  The gauge
field is more interesting.  Here, the periodic
variable $\theta$ will be the Wilson line of the
gauge field around $X_1$,
\be
\theta=M_s\int A_1.
\label{eq:thetatypeI}
\ee
At large $R$, its decay constant, relative
to the Planck scale, has an extra factor of $g$ compared to the closed
string theories discussed previously:
\beq
{f_A^2 \over M_p^2} = {g \over M_s^2 R^2}.
\eeq

On the face of it it seems possible to obtain a large $f$ in one of two
regimes: $R$ comparable to the string scale and $g$ large or
alternatively small $g$ and $R\ll l_s$.  We will consider each of these
regimes in turn but in either case we will see that there is a dual
theory which provides a more appropriate description,
and that $f_A$ is smaller than the Planck scale.

First consider the case of small $g$.  Since $R$ is smaller than the
string scale we should study the T-dual Type I$^\prime$
theory\cite{Polchinski:1995df}.  Type I$'$ is equivalent to Type IIA string
theory on $S^1/\Omega\circ {\bf Z_2}$ with radius $R'=\alpha^\prime/R$.  At
the two fixed points of ${\bf Z_2}$ there are two orientifold O8 planes, with
16 D8 branes in between.  The locations of the D8 branes correspond to the
Wilson line of the type I theory.
To compute $f_A$ in this situation we need to recall
that the low  energy SUGRA action is that of massive IIA supergravity with the
usual form of the  Einstein-Hilbert action. In addition to the SUGRA
contribution to the action we  need to include the worldvolume action coming
from the D8-branes.  This is given by SYM theory in nine dimensions.
The only part of the action that is relevant to us is that of the transverse
fluctuations of the D8-branes. Let $X_9$ be the coordinate describing
these fluctuations.  Then the Wilson line axion is $X_9/R'$.
The coordinate $X_9$ has periodicity $2\pi R'$ and the
axion now has kinetic term,
\be
M_s^7 \int d^{9}x \sqrt{-g} e^{-\phi} (\partial \theta)^2
(\frac{R'}{l_s})^2,
\ee
thus in the type I$'$ theory,
\be
\frac{f_A^2}{M_p^2}=g' M_s R'.
\ee
The coupling in the $T$-dual theory is larger than the original coupling by a
factor of $(R'/R)^{1/2}$.  So we recover the result of equation 8
in the Type I$^\prime$ theory.  It would then appear that if we
take $R^\prime$ large, we can obtain a large $f_A$. We simply
require that $g^\prime > {1 \over R^\prime M_s}$.

But Polchinski and Witten,
in~\cite{Polchinski:1995df} noted that this region of the Type
I$^\prime$ theory is not weakly coupled; in fact, $R^\prime =
{l_s\over g^\prime}$ is a limiting length in the theory.  This
follows from considering Type I-$O(32)$ duality (Polchinski and
Witten actually viewed this as a test of this duality).  In terms
of the original Type I variables,
$g_h = {1 \over g_I}$, while the radius goes over to
$R_h^2= {1 \over g_I} R_I^2$.  So in the dual heterotic theory,
$R_h \sim M_s^{-1}$.  But this is the limiting, self-dual radius
of the O(32) theory.  Thus $f_A$ cannot be parameterically larger
than $M_p$.

Polchinski and Witten explained how this limiting radius shows up
in the Type I$^\prime$ theory:  at these
values of $g' R'$ perturbation theory of type I$'$ string theory breaks down.
The dilaton has a non-trivial profile.  This can
be understood as arising from the non-local cancellation
of tadpoles.  This profile is linear to leading order in $g' R'$,
as expected in this quasi-one dimensional geometry; perturbation
theory breaks down when
$g' R'$ is of order 1.


Now consider the other possibility that the compactification radius is of
order the string scale and the string coupling is large.  Here the theory is
described by heterotic $SO(32)$ theory, at weak coupling.  Under
this transformation, $g_h = {1 \over g_I}$, while the radius goes over to
$R_h^2= {1 \over g_I} R_I^2$.  The radius $R_h$ is small and we must
$T$-dualize to weakly coupled heterotic $E_8\times E_8$ with large radius.
Following this path of dualities and taking into account their effect on
$\theta$ as defined in (\ref{eq:thetatypeI}) we again
find that $f$ is bounded above
by $M_p$.

\subsection{Type IIA}

It is instructive to first consider the case of Type IIA compactified on
a 6 torus.  Because of the high degree of supersymmetry, all extreme
regions of the theory are equivalent to some large radius, weak coupling
description.    The way in which this works
can be rather intricate.  The Type IIA string has a Ramond-Ramond one form gauge
potential, $C_1$ and $\theta$ is the Wilson line of this gauge field around
one of the compact toroidal directions, $\Sigma_1$
\be
\theta=M_s\int_{\Sigma_1} C_1.
\ee

To obtain the kinetic term for $\theta$ recall that the Ramond-Ramond field
enters the 10 dimensional supergravity action as,
\be
S = M_s^8\int d^{10}x\sqrt{-g} |F_2|^2,
\label{eqn:RR}
\ee
where $F_2$ is the field strength of $C_1$.  Compactifying on a torus of
volume $V_5 R$ where $R$ is the size of $\Sigma_1$ results in a decay constant
for $\theta$ of,
\be
f_A^2=\frac{M_s^6 V_5}{R}.
\ee
In order to compare to the four dimensional Planck scale recall the form of
the Einstein-Hilbert action in string frame,
\be
S = M_s^8\int d^{10}x\sqrt{-g}e^{-2\phi} R.
\ee
Note the appearance of the string coupling in the action in contrast to
(\ref{eqn:RR}).  Consequently $M_p^2=M_s^8 g^{-2} V_5 R$.  Measured in units
of the four dimensional Planck scale,
\be
\frac{f_A^2}{M_p^2}=\frac{g^2}{(M_s R)^2}.
\label{eqn:foverm}
\ee
We can attempt to obtain a large decay constant by taking $R\sim l_s$
and $g\gg 1$.  For simplicity we will assume that $V_5$ is large compared to
the string scale.  When $g\gg 1$ one might try to obtain a
description of
the physics in the 11 dimensional supergravity dual.
Rewriting (\ref{eqn:foverm}) in 11
dimensional units one sees that $(f_A/M_p)^2 = (R_{11}/R)^2$ where
$R_{11}=g^{2/3}l_{11}$ is the (large) radius of the 11 dimension.  Naively, it
would seem possible to realize large $f_A$.  However,
\be
R = l_{11} g^{-1/3} = l_{11} \frac{l_{11}}{R_{11}}\ll l_{11}.
\ee
so a better description of this limit is in fact weakly coupled type IIA
string theory on a torus whose sizes are given by $V_5$ and $R_{11}$ both of
which are large in string units, with small string coupling, $g_s^\prime =
R/\ell_{11}$.
In this description, the coupling is weak and all radii large, and
$f_A$ is small.  In the more general case where one or more of the
radii of $V_5$ is small the conclusion applies although the
details vary.

Using the results of \cite{wittenusc} we can understand the
results for toroidal compactification in a much more general way.
Witten has shown that the space of all compactifications to four
dimensions with $32$ supercharges is a connected space.
Furthermore, every asymptotic direction of this space can be
mapped by duality transformations into either Type II string
theory with weak coupling and radii larger than the string scale,
or 11D SUGRA with radii larger than the Planck scale.  In these
regimes it is easy to see that one never gets axion decay
constants larger than the four dimensional Planck scale. The case
of $16$ supercharges is more subtle and we have not studied it
completely.  The moduli space has disconnected components, only
one of which is connected to weakly coupled heterotic strings. The
results of \cite{tbm} show that a result similar to that of
\cite{wittenusc} is obtained, but there are a variety of
asymptotic regions where fully controlled calculations are not
possible\footnote{These were dubbed {\it sprantiloid} regions by
L. Motl, to emphasize that we could name them, but not understand

them.}.  The simplest of these is F-theory on K3 manifolds.  They
are described by string theory with weak coupling on a large
compact space where the coupling varies and sometimes becomes
strong.   We have seen no evidence for large axion decay constants
in these regimes, but have no definitive proof that they cannot be
found.

\section{Theories with Less Supersymmetry}

With less than 16 supersymmetries, the structure of the moduli space is not
well understood.  There is no simple argument that all extreme
regions are equivalent to large radius, weakly coupled theories.  So for such
theories one might hope to find parameterically large decay constants.
In this section we consider several examples, and find that in an appropriate
sense, decay constants cannot be larger than $M_p$.  Our search cannot be
considered exhaustive, but we suspect it is indicative of the general
situation.

\subsection{N=0 Supersymmetry}

The space of (classical) string states with zero supersymmetry is not well
explored.  Toroidal compactifications of Type II strings yield theories with
$32$ supercharges; of the heterotic string, $16$ super
charges. We can obtain
a theory with no supersymmetries by compactifying the Type II theory on a
torus with Scherk-Schwarz boundary conditions\cite{rohm}.  These theories have
a one loop instability, and may not really exist, but we can examine them
using weak coupling string theory. In this case
a naive treatment yields large $f_A/M_p$ for small radii. But
$T$-duality again implies that the correct description of the
theory has a small $f_A$ (in general, the small radius theory is
equivalent to some other non-supersymmetric string theory). Little
is known about the strong coupling limits of these theories, so it
is difficult to make definitive statements as in the
supersymmetric case, but most likely the strong coupling limit, if
it makes sense, is again equivalent to some weakly coupled theory.
In any case, one cannot obtain a large decay constant in any
controlled approximation.

\subsection{Eight or Four Supersymmetries}

With four or eight supersymmetries, much more is known about the theories, but
there are richer possibilities than in the more supersymmetric cases.
The previous examples
illustrated that in compactifications in which small radius is
equivalent to another theory at large radius, one cannot obtain
parametrically large decay constants.  So it is necessary to
consider models where such an equivalence does not hold in such a
simple way.  An example is provided by IIA theory compactified on
a Calabi-Yau space near a conifold singularity.  Near the conifold
point in moduli space, there is a singular region in the geometry
which is topologically $S^2 \times S^3$, with each sphere
shrinking to zero size.

Consider the (pseudo)-scalar field arising from the three form
Ramond-Ramond field $C_{MNO}$:
\beq
\theta=M_s^3\int_{S_3} C_{IJK} d\Sigma^{IJK}
\eeq
where $\Sigma_{IJK}$ is the volume form of the
three sphere. This has unit periodicity in string units, but its
decay constant is of order: \beq {f_A^2 \over M_p^2}= g^2{1 \over
(M_s^3 V_{S^3})^2}. \label{c3f} \eeq The extra factor of $g$
arises because $C$ is a Ramond-Ramond field.  Here there is no
simple dual description of the theory for small $V_{S^3}$, so one
might hope to obtain a large $f_A$.

New light states, however, do appear in the theory
as one shrinks $V_{S^3}$.  These
arise from $D4$ branes wrapped on the $S^3$.  These states
couple magnetically to the $C$ field, and alter the axion effective
action.
These are strings of tension
\beq
T_{D4}= {1 \over g} M_s^5 V_{S^3}.
\label{d4tension}
\eeq
In the limit that $f_A/M_p$ is large, these states become light
compared to the string scale.
Thus it is not possible to simultaneously keep the wrapped $D4$ brane
states heavy and have a decay constant larger than the Planck scale.

In addition, there are D2 brane instantons, whose action is
precisely $M_p /f_A$ in the region where $f_A / M_p$ is large. An
$n$ instanton contribution to the action would be of order $e^{- n
{M_p \over f_A}} e^{in\theta}$, where $\theta$ is the
dimensionless axion field with period $2\pi$. Thus, the instanton
series would effectively terminate only at $n \sim f_A / M_P$ and
the action for canonical fields is rapidly varying.   In this
highly supersymmetric case, there is no axion potential, but we
will see below that a similar mechanism produces rapidly varying
potentials in models with only four supercharges.

\subsection{Type IIB}

Now we consider the Type IIB theory in singular
Calabi-Yau compactfications.  In particular we will consider geometric
singularities due to a collapsing 2 cycle, $S^2$. Examples include a
conifold point, and an $A-D-E$ type singularity in the case of a $K3$
fibration.

Recall that the bosonic part of the type IIB supergravity action includes the
terms:
\be
S_{bos}=\int d^{10}x \sqrt{-g}e^{-2\phi}(R-\frac{1}{2}|H_3|^2)- M_s^8
\int\sqrt{-g} \frac{1}{2}|F_3|^2 + \dots,
\ee
where $H_3=d B_2$ and $F_3=dC_2$ and $C_0$ has been set to zero.  Now one can
hope to obtain a large $f_A$ as $V_{S^2}\rightarrow 0$.  There are two 2 form
gauge fields from which one can obtain an axion: the RR sector two-form field
$C_2$ and the NS-NS sector, $B_2$.  First consider $B_2$,
\beq
\theta=M_s^2\int_{S_2} B_{IJ} d\Sigma^{IJ}.
\eeq

There are additional light states in
this limit coming from wrapping three branes on $S^2$.
But they do not couple directly to the two form field, so one
might hope that they would not interfere with obtaining a small
$f_A$.

Dimensional reduction of the kinetic term for $B_2$ results in a decay
constant for $\theta$ of,
\be
\frac{f_A^2}{M_p^2}=\frac{1}{(V_{S^2}M_s^2)^2}.
\ee
Again, it seems that it is possible to obtain a large $f_A$ provided the $S^2$
is small in string units.  In string perturbation theory,
the action is independent of $\theta$; the leading effects which violate
the continuous shift symmetry are worldsheet instantons.
Their contribution to the effective action is a function of
$u={\rm exp}(M_s^2\int_{S^2} \sqrt{-g}+i\theta)$.  In the limit required for a
large $f_A$ the large $n$ sector instantons are not suppressed.  The effective
action for the canonical axion field varies on scales $f_A/n \sim M_p$.

Similarly one can carry out the analysis for $C_2$, defining
\be
\theta=M_s^2 \int_{S^2}C_{IJ}d\Sigma^{IJ}.
\ee
This field has a decay constant
\be
\frac{f_A^2}{M_p^2}=\frac{g^2}{(V_{S^2}M_s^2)^2}.
\ee
Now it seems possible to obtain a large $f_A$ whilst keeping $V_{S^2}$ large in
string units, by going to strong coupling.  Alternatively we can keep $g$
perturbative and consider shrinking the $S^2$ to below the string scale.
Either case suffers from the same problem of instanton corrections mentioned
above.  This time the instanton corrections come from D-string worldsheet
instantons and are a function of $u={\rm exp}(M_s^2/g \int_{S^2}
\sqrt{-g}+i\theta)$.  The extra factor of $1/g$ comes from the fact that the
D-string couples to the RR sector.  Again, the large $n$ instantons contribute
and result in the decay constant being no larger than $M_p$.  The fact that
the analyses for $B_2$ and $C_2$ result in the same conclusion is to be
expected due to the $SL(2,{\bf Z})$ symmetry of type IIB string theory.

\section{M Theory}

We now wish to consider the strong coupling
limit of some of the string theories considered earlier.  We will start with
the $E_8 \times E_8$ heterotic string.  The low energy limit of this string
theory at large string coupling is given by supergravity on $R^{1,9}\times
S^1/{\bf Z^2}$ where the size of the $S^1$ is given by the string coupling.  The
${\bf Z_2}$ action has two fixed points.  At each of these there is a 10 dimensional
plane with an $E_8$ gauge theory living on it.  The ${\bf Z_2}$ projection breaks
half of the original 32 supercharges of 11 dimensional supergravity.

The supergravity sector of the low energy effective action of heterotic
M-theory is that of 11 dimensional supergravity whose bosonic sector is given
by
\be
S_{11}=M_{11}^9\int d^{11}x
\sqrt{-g}(R-\frac{1}{2}F_4^2)-\frac{1}{6}C_3\wedge F_4\wedge F_4.
\ee
The final, Chern-Simons, term is necessary for anomaly cancellation
and SUSY.  For the
action to be ${\bf Z_2}$ invariant the gauge field $C_3$ must be odd under the
${\bf Z_2}$
projection.  Consequently, the $C_3$ field must have a component along the 11
direction.

We will consider heterotic M-theory on a Calabi-Yau in order to get a four
dimensional minimally supersymmetric theory.  In particular we will
consider the point in the moduli space of the Calabi-Yau where a two cycle,
$\Sigma_2$, shrinks to zero size.  The axion is the $C_3$ field with two
directions on $\Sigma_2$ and the other along the 11 direction,
\be
\theta=M_{11}^3 \int_{\Sigma_2 \times S^1/{\bf Z_2}}C_{ij11}.
\ee
At length scales larger than $R_{11}$ we have an effective four dimensional
theory containing 4 dimensional supergravity, non-abelian gauge theory and the
axion.

Dimensionally reducing the kinetic term for $C_3$, the four dimensional
effective action has a kinetic term for $\theta$ given by,
\be
S =  \frac{V_{CY}R_{11}}{M_{11}^3 R_{\Sigma_2}^4}\int d^4
x\sqrt{-g_4}(\partial \theta)^2.
\ee
In units of the four dimensional Planck scale, the axion's decay
constant is,
\be
\frac{f_A^2}{M_P^2}=\frac{1}{M_{11}^6 R_{11}^2 R_{\Sigma^2}^4}.
\ee
Naively it seems plausible that by shrinking the two cycle one may reach the
regime where $f_A \gg M_P$.  In fact, one needs $R_{\Sigma_2}\ll l_{11}$ while
keeping $R_{11}$ large in 11 dimensional units as required by the large
expectation value of the heterotic dilaton.

However, in this limit the classical supergravity analysis is incomplete since
the shrinking two cycle leads to a geometric singularity.  One must worry
about what physics resolves this singularity.  In particular, one may
worry about new light states coupled to $\theta$ appearing at the
singularity.  This would alter the form of the $\theta$ potential.

Even without worrying about such states, however, we can see that, once again,
high harmonics are likely to be important.
Consider the effect of membrane
instantons.  These give contributions to the action governed by powers of $u$,
where
\be
u=e^{-M_{11}^3\int_{\Sigma_2 \times S^1/{\bf Z_2}}\sqrt{-g}+i \int_{\Sigma_2\times
    S^1/{\bf Z_2}} C_{ij11}}.
\label{membraneinstantons}
\ee
In particular, they are not small in the limit where $f_A$ is large.

In the weak coupling limit, the resolution of the singularities
is understood.
The lowest order, classical
supergravity analysis is modified by perturbative corrections in $\alpha'$ as
well as worldsheet instantons.  Witten showed~\cite{Witten:1999fq} that the
only non-zero contributions come from lowest order in $\alpha'$ and from
worldsheet instantons.  These effects alone resolve the singularity with no
new massless states in the low energy description.   However the inclusion of
the worldsheet instanton effects modify the axion effective action.
In the weak coupling limit, the membrane instantons are just the world sheet
instantons.  World sheet instanton effects
are functions of $u={\rm exp}(-T\int_{\Sigma_2}\sqrt{-g}+i \int_{\Sigma_2}B)$.   This
is precisely the same quantity as in eqn. \ref{membraneinstantons}, translated
into the weak coupling language.

So, once again, we see that the dominant harmonics have
$n$ up to $n\sim f_A/M_p$.  So the Fourier series for the
effective action will converge slowly, and will be a rapidly
varying function of $\theta$; the action for the canonical
field will vary on the scale $M_p$.


\subsection{$G_2$ Example}

Finally we will consider an example of 11D SUGRA compactification
on a manifold with $G_2$ holonomy.  An example of interest,
closely related to the conifold of string theory, is that of a
$G_2$ manifold which is asymptotically a cone over a six
dimensional manifold which is $S^3 \times S^3$.  Topologically the
space is $R^4 \times S^3$.  There is a vanishing supersymmetric
cycle which is a three sphere, $X$.  The axion is now, \be \theta
=M_{11}^3\int_{X} C_3, \ee where $C_3$ is again the three form of
11 dimensional supergravity.

$C_3$ appears in the supergravity action as,
\be
M_{11}^9\int d^{11}x \sqrt{-g} |F_4|^2
\ee
from here one sees that the decay constant in Planck units is,
\be
\frac{f_A^2}{M_P^2}=\frac{1}{(V_X M_{11}^3)^2}.
\ee
In order to get large $f_A$ we need the shrinking three cycle to be small
compared to the 11 dimensional scale, $l_{11}$.  As before, the supergravity
analysis is corrected by instanton effects.  In this case the instantons come
from M2 branes wrapping $X$.  Their contribution to the action will be a
rational function of $u$ where,
\be
u=e^{-M_{11}^3\int_{X}\sqrt{-g} + i \int_{X} C}.
\ee
These membrane instanton effects are not suppressed in the limit of small
volume of $X$.

The large $n$ instanton contributions are no longer suppressed and
the axion decay constant is restricted to be $f_A/n$.  The instanton
effects mean that the effective action is once again
dominated by high harmonics, so that the potential varies on
the Planck scale.

A similar analysis applies to any $G2$ singularity where some
number of three cycles shrinks to zero.
\section{Conclusions}

Axion like fields, with decay constants larger than the Planck
scale, could give rise to models of natural inflation, and if the
decay constant were extremely large, might form the basis for an
anthropic explanation of the value of the cosmological constant.
One can only study the plausibility of such scenarios in a
framework where Planck scale physics is under control.  This is
the case in certain extreme regions of the moduli space of
string/M theory, and, as far as we are aware, only in this arena.

We have examined a variety of regions in moduli space where a
large decay constant could arise, but have found no consistent
scenario of this type.  In each case the required axion field
arises from the components of a p-form gauge potential along a p-cycle in the
limit that the cycle shrinks to zero size.  There are always light states
and/or low action instantons, which arise in this limit, and give rise to
rapidly varying contributions to the axion potential (in those cases where a
potential is allowed by supersymmetry) which effectively rescale the decay
constant to the Planck scale.

Another way that one might imagine that string moduli might serve
the purposes of natural inflation is in the noncompact regions of
moduli space.  However, in those regions, general arguments show
that the potential falls so rapidly that no solutions with
accelerated expansion can be found.  The motion of the moduli in
such noncompact regions is dominated by kinetic energy, rather
than friction.

Although our arguments sound rather general, we have certainly not
explored all possible singular limits of M-theory, so it would be
premature to claim that our results are a no-go theorem.  In our
opinion they seem suggestive of such a theorem, and it seems worth
while to explore further, in search of a counterexample, or a
proof.


\noindent
{\bf Acknowledgements:}

\noindent
We would like to thank W. Fischler for interesting conversations
and for exhibiting extreme skepticism when we thought we had found
examples with large decay constant.
We also thank Scott Thomas for prodding us to consider further
examples.  We thank  P. Creminelli for pointing out a misstatement about Type
I theories.  This work supported in part by the U.S.
Department of Energy.



\begin{thebibliography}{99}
\singlespaced


\bibitem{gaillard}
P.~Binetruy and M.~K.~Gaillard,
Phys.\ Rev.\ D {\bf 34}, 3069 (1986).

\bibitem{banks}
T.~Banks,
arXiv:hep-th/9906126.

\bibitem{thomas}

S.~Thomas,
Phys.\ Lett.\ B {\bf 351}, 424 (1995)
[arXiv:hep-th/9503113].

\bibitem{natural}
K.~Freese, J.~A.~Frieman and A.~V.~Olinto,
Phys.\ Rev.\ Lett.\  {\bf 65}, 3233 (1990).


\bibitem{hw}
P.~Horava and E.~Witten,
Nucl.\ Phys.\ B {\bf 460}, 506 (1996)
[arXiv:hep-th/9510209].




\bibitem{nimaetal}
N.~Arkani-Hamed, H.~C.~Cheng, P.~Creminelli and L.~Randall,
arXiv:hep-th/0301218.


\bibitem{Kaplan:2003aj}
D.~E.~Kaplan and N.~J.~Weiner,
arXiv:hep-ph/0302014.



\bibitem{Witten:1999fq}
E.~Witten,
JHEP {\bf 0002}, 025 (2000)
[arXiv:hep-th/9909229].


\bibitem{rohm}
R.~Rohm,
Nucl.\ Phys.\ B {\bf 237}, 553 (1984).


\bibitem{ginsparg}
P.~Ginsparg,
Phys.\ Rev.\ D {\bf 35}, 648 (1987).

\bibitem{tbm}
L.~Motl and T.~Banks,
JHEP {\bf 9905}, 015 (1999)
[arXiv:hep-th/9904008].



\bibitem{anthropiclambda}
S.~Weinberg,
Rev.\ Mod.\ Phys.\  {\bf 61}, 1 (1989).




\bibitem{bankslight}
T.~Banks,
Phys.\ Rev.\ Lett.\  {\bf 52}, 1461 (1984).

\bibitem{villenkin}
J.~Garriga and A.~Vilenkin,
Phys.\ Rev.\ D {\bf 64}, 023517 (2001)
[arXiv:hep-th/0011262];
J.~F.~Donoghue,
JHEP {\bf 0008}, 022 (2000)
[arXiv:hep-ph/0006088].

\bibitem{sti}
M.~Dine,
arXiv:hep-th/0107259.


\bibitem{wittenusc}
E.~Witten,
Nucl.\ Phys.\ B {\bf 443}, 85 (1995)
[arXiv:hep-th/9503124].


\bibitem{Giveon:1994fu}
A.~Giveon, M.~Porrati and E.~Rabinovici,
Phys.\ Rept.\  {\bf 244}, 77 (1994)

[arXiv:hep-th/9401139].

\bibitem{Polchinski:1995df}
J.~Polchinski and E.~Witten,
Nucl.\ Phys.\ B {\bf 460}, 525 (1996)
[arXiv:hep-th/9510169].

\end{thebibliography}
\end{document}